# The Distributed Genetic Algorithm Revisited [*]

Theodore C. Belding
Division of Computer Science and Engineering
University of Michigan
Ann Arbor, MI 48109-2122 USA
E-mail: Ted.Belding@umich.edu

## Abstract

This paper extends previous work done by Tanese on the distributed genetic algorithm (DGA). Tanese found that the DGA outperformed the canonical serial genetic algorithm (CGA) on a class of difficult, randomly-generated Walsh polynomials. This left open the question of whether the DGA would have similar success on functions that were more amenable to optimization by the CGA. In this work, experiments were done to compare the DGA's performance on the Royal Road class of fitness functions to that of the CGA. Besides achieving superlinear speedup on KSR parallel computers, the DGA again outperformed the CGA on the functions $R3$ and $R4$ with regard to the metrics of best fitness, average fitness, and number of times the optimum was reached. Its performance on $R1$ and $R2$ was comparable to that of the CGA. The effect of varying the DGA's migration parameters was also investigated. The results of the experiments are presented and discussed, and suggestions for future research are made.

## 1 INTRODUCTION

Reiko Tanese (1987, 1989a, 1989b) proposed the distributed genetic algorithm (DGA) as a way of efficiently parallelizing the canonical genetic algorithm (CGA) on hypercube computers and other MIMD parallel computers. The global population is divided into several subpopulations, one per processor. Each processor runs the CGA independently on its subpopulation. The only inter-processor communication occurs during the migration phase, which takes place at a regular interval: A fixed proportion of each subpopulation is selected and sent to another subpopulation. In return, the same number of migrants are received from some other subpopulation and replace individuals selected according to some criteria. This migration can occur either asynchronously or after all of the processors have been synchronized. Each processor then resumes running the CGA as before, until the next migration phase.

Tanese found that the DGA achieved near-linear speedup on a 64-processor NCUBE/six hypercube computer. She then examined whether parallelizing the CGA in this manner hurt its performance when optimizing a class of Walsh polynomials now known as the Tanese functions. The DGA generally found fitter individuals than the CGA. In addittition, the DGA achieved levels of average fitness comparable to the CGA when the migration interval $i$ and the migration rate $r$ were set such that about 1% of the subpopulation migrated per generation, e.g., 20% of the subpopulation migrating every 20 generations (or $i = 20$ and $r = 0.2$). Surprisingly, hillclimbing outperformed both the CGA and DGA on these functions; furthermore, the partitioned genetic algorithm, a DGA without migration, found fitter individuals than the CGA, even with a subpopulation size as small as 2 individuals.

Forrest and Mitchell (1991) examined the CGA's performance on the Tanese functions. They concluded that it performed poorly mainly because the effective order of the schemata in the functions was much higher than intended. The lack of low-order schemata also played a role, as did the long defining length of those schemata that were present. Tanese chose to work with these functions because they were constructed using partitions which bore a certain resemblance to the schemata processed by genetic algorithms (GAs). In

---

retrospect, the Tanese functions "have more to do with parity than with schemata" because of the way these functions are defined (Q. F. Stout, personal communication). The question of the DGA's efficacy on other, less pathological functions was left open.

This paper contributes towards filling this gap by evaluating the performance of the DGA relative to the CGA on a class of fitness functions called the Royal Road functions: $R1$ and $R2$ (Mitchell et al. 1992; Forrest & Mitchell 1993), $R3$ (Mitchell & Holland 1993), and $R4$ (Mitchell et al. 1994). Though highly contrived, these functions have a fixed number of predetermined schemata, allowing researchers to study GA performance over time. Due to space constraints, they will not be described here. Besides achieving superlinear speedup on KSR parallel computers, the DGA is found to consistently outperform the CGA on the functions $R3$ and $R4$; it achieves results comparable to the CGA on $R1$ and $R2$. The following section briefly surveys related work on distributed genetic algorithms and topics in evolutionary biology. The results of the experiments are then presented and discussed. The paper concludes with a summary and suggestions for future research.

## 2 RELATED WORK

Since their inception, it has been clear that GAs are inherently parallel algorithms. Beginning in 1987, a wide variety of parallel implementations have appeared in the literature. This paper focuses on one of these, the island model.

Island-model GAs derive their name from population genetics, where the term denotes a population split into many semi-isolated subpopulations, like an archipelago of islands. These GAs are also called medium- or coarse-grained parallel GAs, or distributed GAs. Besides Tanese's DGA (Tanese 1987, 1989a, 1989b), island-model GAs have been investigated by Pettey et al. (1987), Cohoon et al. (1987), and Sumida (1990), among others. Whitley (1993) has investigated the DGA using a formal model.

Tanese suggested that the success of the DGA relative to the CGA on certain fitness functions might bear some relation to Sewall Wright's shifting balance theory in evolutionary biology (Wright 1932; Crow 1991), which predicts that a loosely-connected network of small subpopulations, or demes, may sometimes evolve more rapidly towards a global fitness peak than a single large population. Wright hoped to explain how populations in nature are able to escape local optima and discover novel gene complexes. The process works as follows: The subpopulations drift randomly around a local fitness peak. If one of the subpopulations happens to drift across the intervening fitness valley into the influence of a higher fitness peak, it is pulled up this new peak by intrademic natural selection. This subpopulation produces an excess of offspring, due to its high average fitness, which then emigrate to the other subpopulations, spreading the newly-discovered gene complex throughout the population. The process can also occur in a continuous population of small overlapping neighborhoods (Rouhani & Barton 1987). Theoretical studies have shown that a surprisingly small amount of migration suffices to disperse an advantageous gene complex throughout the population (Phillips 1993). This research has been supported by laboratory experiments (Wade & Goodnight 1991), but it is unclear whether the right combination of migration, drift, and selection exists in nature for this process to take place. It is also debated whether local fitness peaks are common in the high-dimensional fitness spaces of organisms (Turner 1987).

Cohoon et al. (1987) argued that the DGA's success was due to another phenomenon, Eldredge and Gould's theory of punctuated equilibrium. This theory states that evolution is characterized by long periods of relative stasis, punctuated by periods of geologically rapid change associated with speciation events (Eldredge & Gould 1972). Originally, Eldredge and Gould saw their theory as a consequence of Mayr's (1963) theory of speciation in small, peripheral populations; more recently, Mayr's theory has been questioned (Coyne 1994; Gould & Eldredge 1993). Cohoon et al. pointed out that the GA also tends towards stasis, or premature convergence. They argued that isolated "species" could be formed by separating the global populations into subpopulations. By injecting individuals from a different species into a subpopulation after it had converged, new building blocks would become available; furthermore, the immigrants would effectively change the fitness landscape within the subpopulation. These two factors together would induce a "speciation event," similar to speciation in a peripheral isolate, which in turn would be accompanied by a period of rapid evolution. The DGA could thus partly avoid the problem of premature convergence. Futuyma (1987) has suggested that rapid evolution may be associated with speciation events because the morphological change that accumulates in a population can only escape being broken up by recombination if the population speciates and becomes reproductively isolated from the remainder of the old species. In the GA, this may translate into an advantage for populations that are divided into discrete subpopulations, and thus are able to follow separate evolutionary trajectories and preserve their diversity.

Holland (1988), Hillis (1988), and Vose and Liepins (1991) have also discussed the apparent prevalence of punctuated equilibrium in the performance of GAs. Holland attributed it to the rapid increase of highly-fit building blocks predicted by the Schema Theorem. When the population's average fitness approached the fitness of the best individuals present, the period of exponential fitness increase would come to an end. Vose and Liepins argued that the phenomenon was due to the presence of unstable attractors within the fitness landscape: As the population approached an attractor, its rate of evolution would decrease. When the population came under the influence of a different attractor due to crossover, another period of exponential fitness increase would commence. It is interesting to note that Wright (1982) argued that the shifting balance theory alone is sufficient to account for the pattern of punctuated equilibrium in the fossil record; Vose and Liepins' paper is essentially an independent, mathematical restatement of Wright's ideas. Hillis also cited transitions between local fitness peaks as one reason for punctuated equilibrium but suggested that epistasis could cause it as well: Punctuated equilibrium could result if a trait with high fitness depended on specific alleles being present at several loci simultaneously.

## 3 SPEEDUP

In trials done on a 2-ring, 64-processor KSR1 computer, the DGA exhibited superlinear speedup for up to 32 processors (Figure 1). Timing was averaged over 200 runs, with a fixed number of evaluations of $R1$ during each run. The processors were synchronized only during the migration phase; the migration interval was set to 5 generations, and 10% of each subpopulation was exchanged during migration. More recently, the run time was examined for a fixed number of evaluations of $R4$ on a 2-ring, 64-processor KSR2. Timings were averaged over 500 runs on 1 processor and on 24 processors, and the program running on 24 processors was synchronized after every generation. Again, superlinear speedup was achieved. The reason for this superlinear speedup is not entirely clear. It may be that the DGA program and data are being subdivided sufficiently that each processor can fit its entire portion of the program into its subcache, instead of the KSR's main cache, thus speeding memory accesses relative to the CGA (Q. F. Stout, personal communication).

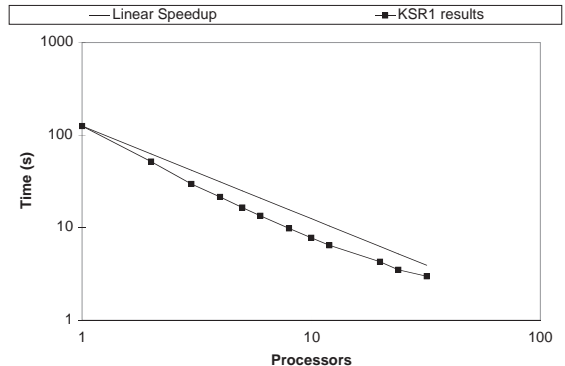

Figure 1: KSR1 speedup

## 4 EXPERIMENTAL RESULTS

As stated earlier, the main objective of this work was to investigate whether Tanese's results were peculiar to the functions she studied, or whether they would also hold for other functions that were more amenable to optimization by the GA. To this end, the performance of the DGA was compared to that of the CGA on the Royal Road fitness functions $R1$, $R2$, $R3$, and $R4$. The effect of varying the DGA's migration interval and migration rate parameters was also studied.

All experiments were done with a population size of 480. The DGA was run on 24 processors, with one subpopulation of 20 individuals allocated to each processor. The entire population was replaced each generation. Each run lasted 500 generations, or 240,480 function evaluations (counting generation 0). The processors were synchronized after each generation, so that accurate statistics could be collected, as well as during migration. In order to collect data on the CGA, the DGA was run on a single processor with a single subpopulation and no migration.

The number of crossover points per pair was selected from a Poisson distribution with a mean of 2.816, and the mutation operator had a probability of 0.005 per bit of flipping each bit. Selection was proportionate to fitness, and sigma scaling (Tanese 1989b; Forrest & Mitchell 1991) was used in all runs: $f' = f - (\overline{f} - c\sigma)$, where $f'$ is the scaled fitness, $f$ is the raw, objective fitness, $\overline{f}$ is the subpopulation's average objective fitness, $c = 2$, and $\sigma$ is the standard deviation of the raw fitness. If $\sigma < 0.0001$, $f'$ was set to 1. If $f' > 1.5$, it was reset to 1.5; the minimum possible scaled fitness was 0. The subpopulation's av-erage fitness was recalculated after scaling, to keep the subpopulation size constant.

The DGA was studied with migration intervals $i$ of 5, 10, 20, 50, 100, and 500 generations. The DGA with $i = 500$ was equivalent to Tanese's partitioned GA, since each run lasted only 500 generations. Migration rates $r$ of 0.1, 0.2, and 0.5 were studied; i.e., $10\%, 20\%$, and $50\%$ of the individuals in each subpopulation were exchanged, respectively. All 15 possible combinations of $i$ and $r$ were studied. In contrast to the hypercube migration used by Tanese, the recipient and donor subpopulations of the migrants were chosen at random for each migration phase. This was implemented by generating a random permutation of the subpopulations, such that each subpopulation received migrants from a single subpopulation other than itself; furthermore, no 2 subpopulations received migrants from the same subpopulation. Emigrants were chosen by simply taking the first $n_{mig}$ individuals in the subpopulation, where $n_{mig} = r \cdot n$. Here $n$ is the size of the subpopulation. The same $n_{mig}$ individuals were then replaced by immigrants. This differs from the scheme used by Tanese, where both the migrants and the individuals they replaced were selected at random. The migration scheme also differs from that in the shifting balance process: There, the number of emigrants depends on a subpopulation's average fitness. In this study, $n_{mig}$ was fixed identically for all of the subpopulations.

The CGA was run 500 times on each function, as was the DGA with each of the possible migration parameter settings. The fitness of the best individual in the entire population and the global average fitness were recorded each generation; the data were then averaged over the 500 runs. In addition, the number of runs in which the global optimum was reached was recorded for each set of parameters, along with the mean number of generations needed to reach the optimum and the standard deviation. The results are presented in Figures 2–13. In order to increase the figures' legibility, the plots of best fitness and average fitness for each of the functions show only the data series for those migration parameters that produced the best results, plus the CGA and the partitioned GA. Note that on $R4$, 4 trials of 500 runs each were conducted for the CGA, as it failed to reach the optimum in the first 2 trials. Only the first of these is plotted in Figures 11 and 12.

## 5  DISCUSSION

The charts show a clear qualitative difference between the results for $R1$ and $R2$ and those for $R3$ and $R4$. On both $R1$ and $R2$ the optimum was almost always

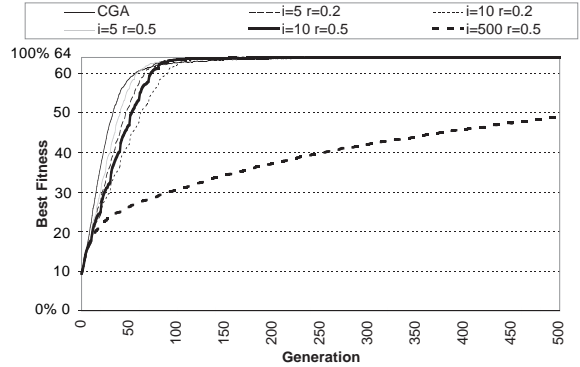

Figure 2: $R1$ best fitness

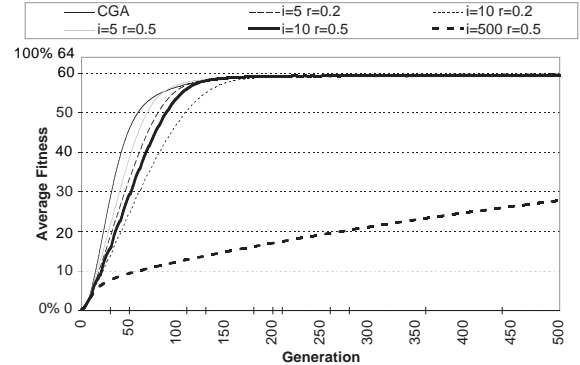

Figure 3: $R1$ average fitness

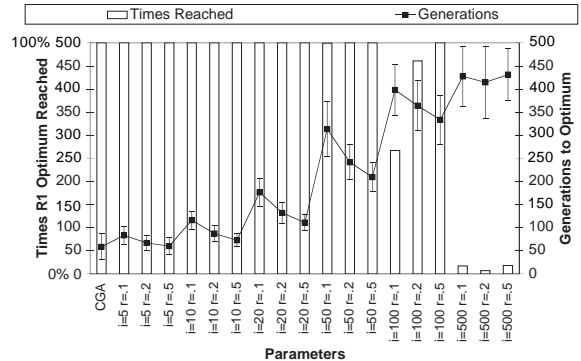

Figure 4: Number of times the optimum was reached on $R1$

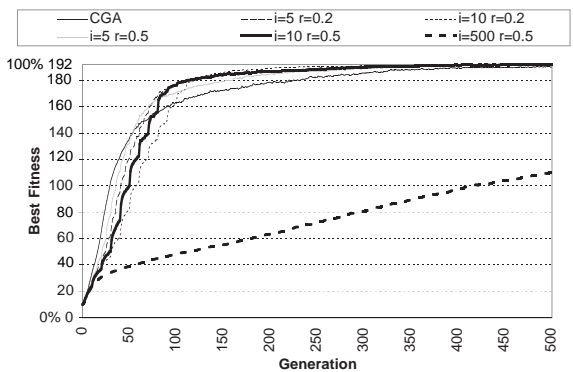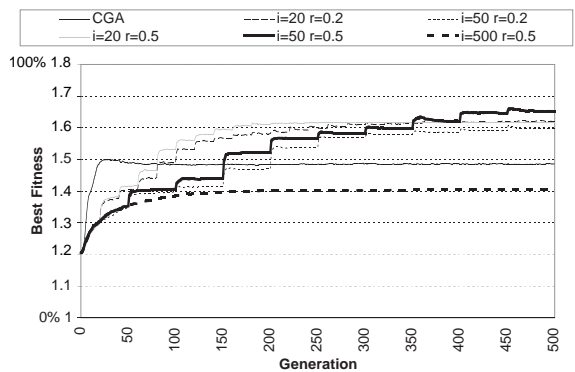

Figure 5: $R2$ best fitness

Figure 8: $R3$ best fitness

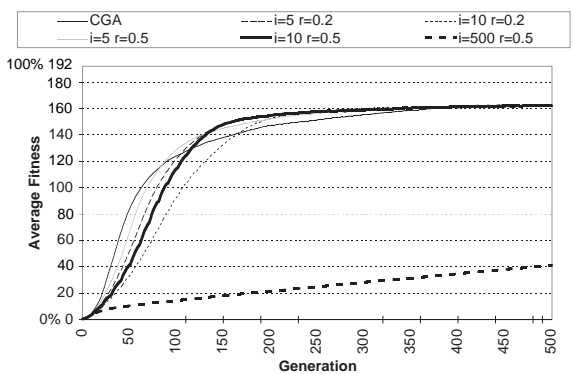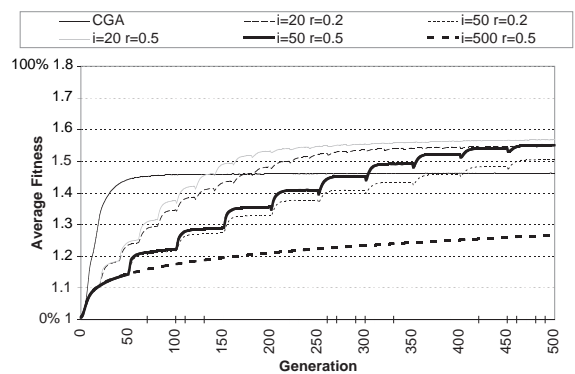

Figure 6: $R2$ average fitness

Figure 9: $R3$ average fitness

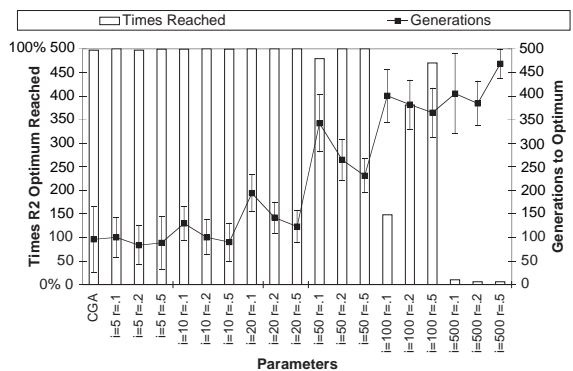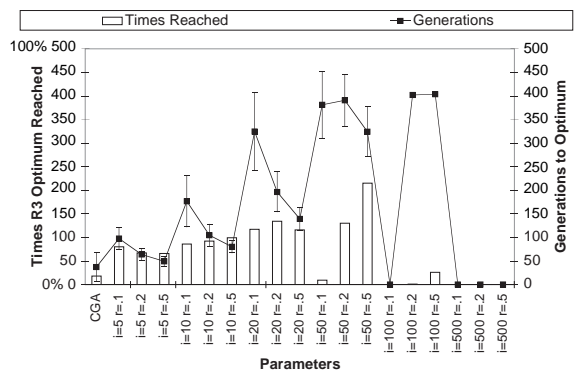

Figure 7: Number of times the optimum was reached on $R2$

Figure 10: Number of times the optimum was reached on $R3$

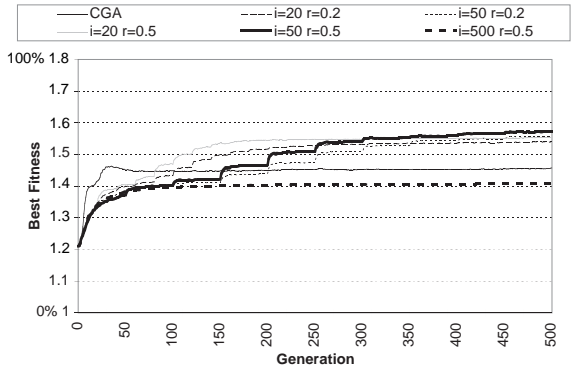

Figure 11: $R4$ best fitness

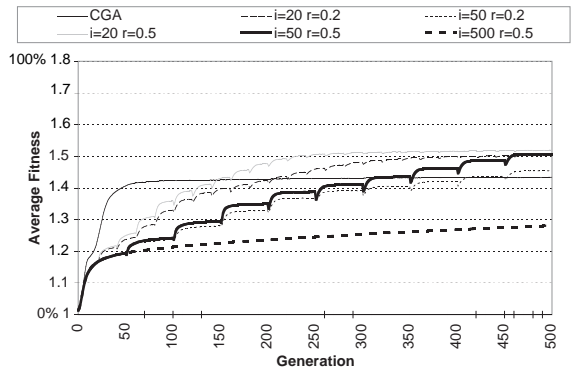

Figure 12: $R4$ average fitness

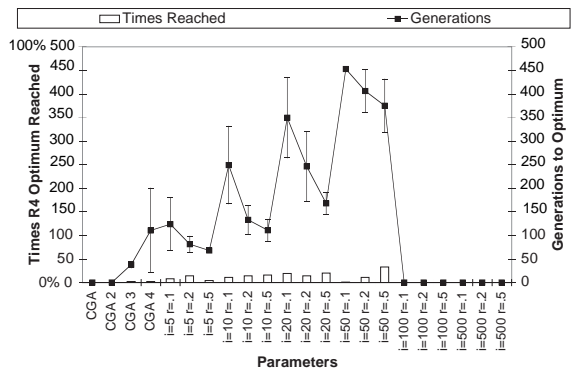

Figure 13: Number of times the optimum was reached on $R4$

reached by both the CGA and the DGA, except for the DGA with $i = 100$ or $i = 500$. Furthermore, the DGA generally performed worse than the CGA on these fitness functions, when the metric of generations needed to find the optimum is taken into account; only the DGA trials with $i = 5$ and $i = 10$ performed comparably to the CGA. In contrast, the CGA rarely found the global optimum for $R3$ and $R4$ (18 out of 500 runs on $R3$ and 1 out of 500 runs on $R4$). The optimum was found most often on these 2 functions by the DGA with $i = 50$ and $r = 0.5$ (215 out of 500 runs on $R3$ and 33 out of 500 runs on $R4$).

Tanese reported that the DGA usually took longer than the CGA to find the optimum, indicating that the DGA was able to search longer because it maintained diversity across the subpopulations and avoided premature convergence longer. This holds true in these results as well. However, the relatively strong performance of the trials with $r = 0.5$ was unexpected; this is a massive amount of migration. In Tanese's work, the DGA performed best with $r = 0.1$. In contrast, trials with $r = 0.1$ generally performed worse on the Royal Road functions than trials with other migration rates, while taking longer to find the optimum. Finally, the partitioned GA ($i = 500$) performed worst of all the parameter settings, on all of the functions; this also differs from Tanese's results, where the partitioned GA often found the fittest individuals.

As Tanese (1989b), Cohoon et al. (1987), and Mühlenbein (1991) have reported, the influx of migrants into the subpopulations causes a temporary dip in the average fitness level, because of the low fitness of many of the hybrids between the new immigrants and the original members of the subpopulation. The levels of both average and best fitness then rise significantly, due to the the discovery of novel, highly-fit building blocks. After a renewed period of exponential increase, these increases taper off, as the average population fitness approaches the level of the best individuals (Holland 1988). The cycle then repeats at the next migration phase. Mühlenbein (1991) pointed out that this phenomenon differs from the type of evolution expected by the shifting balance theory: In Wright's theory, a subpopulation happens upon novel building blocks by genetic drift; these are then spread to the rest of the population by migrants. In the DGA, the building blocks brought by the migrants are often less fit than those created after the migrants interbreed with the native population.

In order to check the possibility that the random migration scheme used in this research was responsible for the differences with Tanese's results, 500 runs were performed on $R4$ for each of the 15 migration parameters and the CGA, using a population size of 512 and 32 subpopulations. Trials were performed using both the random migration topology described above and Tanese's hypercube migration scheme (Tanese 1987, 1989b). The results (not shown, due to lack of space) were essentially identical between the 2 topologies, with some slight variation. Based on these data, it seems unlikely that the random migration used in this research had much effect on the outcome. This was to be expected, since a hypercube or stepping-stone model rapidly approximates a random, island migration topology as the number of its dimensions increases.

It is also significant that certain trials, such as those using the CGA, are characterized by rapid increase in fitness, followed by equally rapid stagnation; others (e.g., the DGA on $R4$ with $i = 50$ and $r = 0.5$) are characterized by slower, yet ultimately more significant increase. In the hope that the best of both worlds could be attained — rapid, yet sustained increase in fitness, 3 trials of 500 runs each were conducted of the DGA with variable migration parameters on the function $R4$ (these results are also not shown). The runs began with $i = 10$ and $r = 0.5$, settings which produce quick, but short-lived increase in fitness. In the first trial, the migration interval $i$ was changed to 20 at the end of generation 140 and changed again to 50 at the end of generation 320; these parameters are characterized by more gradual and more sustained fitness increase. In the second and third trials, the migration interval was changed from 10 to 50 at the end of generation 60, and from 10 to 50 after generation 20, respectively. None of these trials showed significant improvement over either of the two original parameter settings. It seems likely that these disappointing results are due to the fact that the population's diversity is rapidly depleted during the interval when $i = 10$; after the subpopulations have converged, changing the migration interval will produce no benefit.

## 6 CONCLUSIONS

This research project was undertaken to extend Tanese's work on the DGA to the Royal Road class of fitness functions, as a first step towards determining whether her results were specific to the Tanese functions or also held for other functions, on which the CGA performed well. The results raise as many questions as they answer. The DGA achieved superlinear speedup on KSR parallel computers. It also outperformed the CGA on the functions $R3$ and $R4$, using the metrics of best overall fitness, global average fitness, and number of times the optimum was reached; the best results were obtained with migration interval $i = 50$ and migration rate $r = 0.5$. In contrast, the DGA only achieved comparable results to the CGA on the functions $R1$ and $R2$, using migration intervals $i$ of 5 and 10. This may be because these functions are relatively easy for the CGA, as compared to $R3$ and $R4$. The dismal performance of the partitioned GA differed dramatically from Tanese's data. Initial results with variable migration parameters showed no improvement over fixed parameters, probably due to the quick loss of diversity that accompanies a high frequency of migration. Further research is needed in all of these areas. Work is in progress to track the schemata within each subpopulation and to measure population diversity over time, in order to better understand how the DGA sometimes succeeds where the CGA does not. An *a priori* theory of the interaction of the migration parameters and topologies with the fitness function being used would be very valuable. To this end, more realistic fitness functions need to be investigated, once the Royal Road functions are understood. The reasons for the differences between Wright's theory and the DGA results should also be investigated. By doing so, we may not only develop more efficient genetic algorithms, but possibly learn something about evolution in nature as well.

## Acknowledgements

I thank David Ackley, Andy Adams, David Blake, Stephanie Forrest, Greg Gibson, Danny Hillis, John Holland, John Laird, Frances McSparran, Melanie Mitchell, Tim Stanley, the anonymous reviewers, and especially Quentin Stout. This research was performed using KSR1 and KSR2 parallel computers at the University of Michigan's Center for Parallel Computing, which is partially supported under NSF grant CDA-92-14296.